\documentclass[12pt]{article}

\usepackage{physics} 
\usepackage{siunitx} 
\usepackage{enumerate} 
\usepackage{graphicx}

\usepackage{amssymb}

\usepackage{cite}

\begin{document}

\title{Average Trapping time on the 3-dimensional 3-level Sierpinski gasket network with a set of trap nodes}
\author{Zhizhuo Zhang; Bo Wu}
\date{\today}

\maketitle

\begin{abstract}
As a basic dynamic feature on complex networks, the property of random walk has received a lot of attention in recent years. In this paper, we first studied the analytical expression of the mean global first passage time (MGFPT) on the 3-dimensional 3-level Sierpinski gasket network. Based on the self-similar structure of the network, the correlation between the MGFPT and the average trapping time (ATT) is found, and then the analytical expression of the ATT is obtained. Finally, by establishing a joint network model, we further give the standard process of solving the analytical expression of the ATT when there is a set of trap nodes in the network. By illustrating examples and numerical simulations, it can be proved that when the trap node sets are different, the ATT will be quite different, but the the super-linear relationship with the number of iterations will not be changed.
\end{abstract}

\section{Introductiom}

With the vigorous development of the complex network field, the attention caused by the random walk problem on the network is also increasing year by year.\cite{newman2018networks} As the basic dynamic characteristics of the network, the random walk problem has gradually become a research hotspot because it not only plays a key role in studying the topology and dynamic characteristics of the network, but also has an application background in the fields of physics, chemistry, biology and communications, etc.. For example, the random walk model can be used to study and simulate information transmission\cite{chau2011analysis}, data collection\cite{zheng2014data,lee2015towards}, cost of communication search\cite{roberto2009low,lin2008dynamic}, the spread of infectious diseases in organisms\cite{gopalakrishnan2009random}, propagation of porous media\cite{cai2017fractal} and heterogeneous catalysis\cite{vignoles2011pearson}, etc.
Among the many characteristics of random walks, the mean global first passage time (MGFPT) and the average trapping time (ATT) are two types of properties worthy of attention, because they not only reflect the global random walk property of the network, but are more directly related to the efficiency of information collection on the network. It has been proved that the smaller the ATT, the higher the efficiency of information search and diffusion on the network.\cite{montroll1969random}

As a classic network model with self-similar properties, the Laplacian spectrum\cite{bentz2010analytic,rammal1984spectrum}, spanning tree\cite{chang2007spanning}, diffusion\cite{guyer1984diffusion}, unbiased random walk\cite{kozak2002analytic,haynes2008global}, loop-erased random walk\cite{dhar1997distribution} and other properties\cite{chang2008dimer,dhar1997distribution,lin2002electronic} on the Sierpinski gasket network have been systematically studied. However, the main research results obtained so far focus on the most basic 2-dimensional 2-level Sierpinski gasket network, and there are few studies on high-dimensional and high-level Sierpinski gasket networks. In particular, in a network with self-similar properties, MGFPT and ATT have a special correlation, but this feature is not easy to be found in the second-level Sierpinski gasket.\cite{wu2020average} In addition, the topological and dynamic properties of the high-dimensional advanced Sierpinski gasket network are quite different from the two-dimensional secondary network, so it is worthy of further study.

In this paper, we consider the unbiased random walk problem on a 3-dimensional 3-level Sierpinski gasket network, which has not been studied so far. In the section 2, we give the iterative construction method of the network model, which also reflects the global self-similar characteristics of the network. Then, in the section 3, based on the probability generating function method\cite{peng2018moments}, the analytical expression of MGFPT on the network is solved. Furthermore, in the section 4, through the analysis of the path of walker starting from a any node in the network to the trap node, we find the correlation between MGFPT and ATT, and obtain the analytical expression of ATT. In Section 5, we expand the analytical expression of ATT. Based on the matrix algorithm\cite{zhang2021mean}, a standard method for solving the ATT with a trap nodes set in the network is given, and an example is used to illustrate it, which is not considered in the previous research work of random walk.

\section{Structure of the $3$-dimensional $3$-level Sierpinski gasket network}
\begin{figure}[t]
\centering
\includegraphics[scale=0.25]{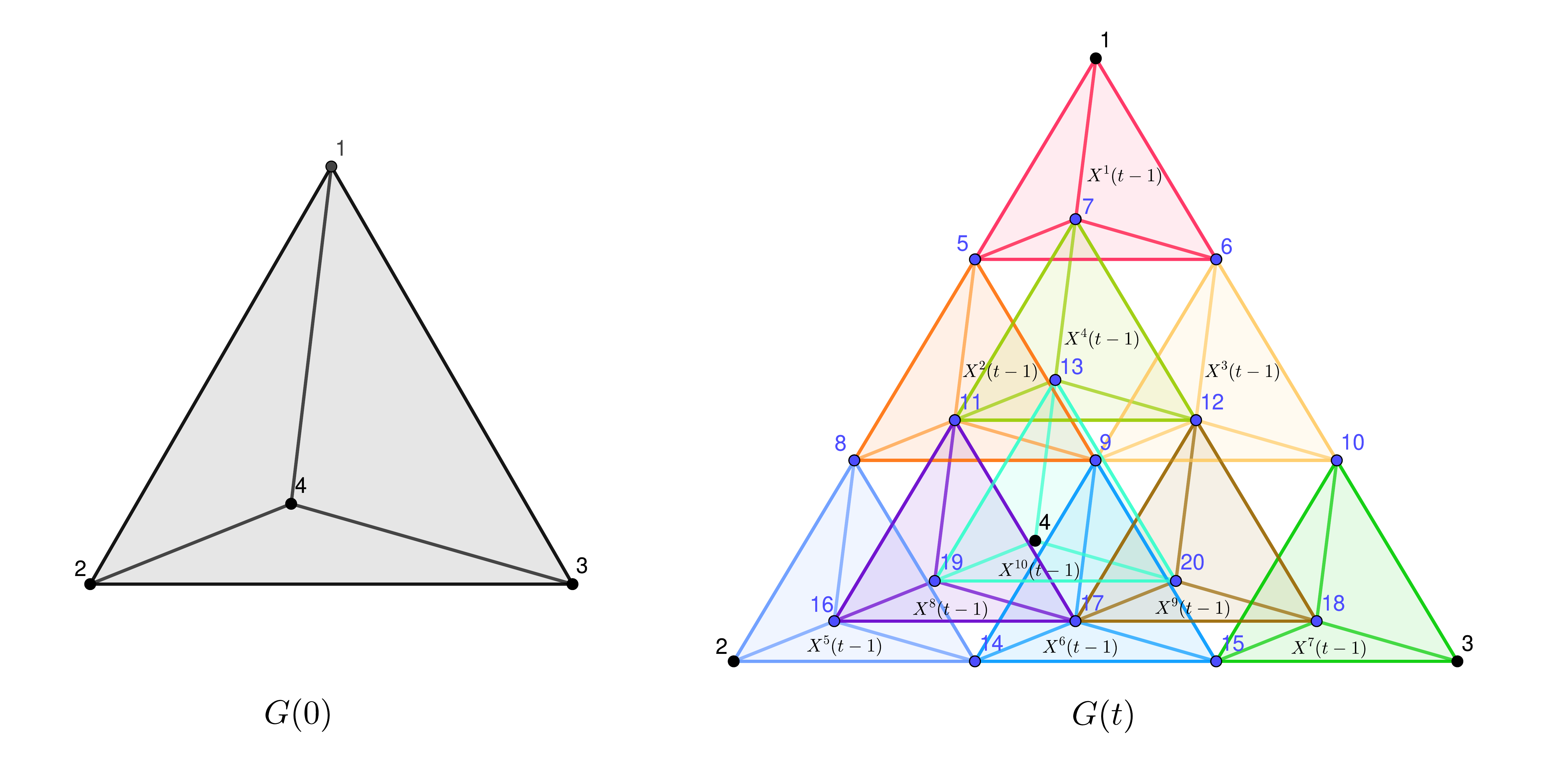}
\caption{The initial network $G(0)$ and the network $G(t)$ after $t$ iterations.}
\label{fig1}
\end{figure}

In this section, we will introduce the construction method and some basic topological properties of the three-dimensional three-level Sierpinski gasket network. Since the $3$-dimensional $3$-level Sierpinski gasket network is constructed by iteration, the number of iterations is denoted as $t$, the initial network is denoted as $G(0)$, and the network after $t$ iterations is denoted as $G(t)$. The initial network $G(0)$ is defined as a regular tetrahedron composed of four nodes, as shown in the left half of Fig \ref{fig1}. The network $G(t)$ can be obtained by splicing $10$ $t-1$ generation networks $G(t-1)$. The combination method is shown in the right half of Fig \ref{fig1}, where $X^i(t-1)$ $(1\le i\le 10)$ represents a local area of the network $G(t)$ and the structure is the same as that of the network $G(t-1)$. 
Obviously, the network $G(t)$ can be regarded as the aggregation network obtained by splicing the local structure $X^i(k)$ $(0\le k\le t-1)$, and the set of all $X^i(k)$ in $G(t)$ is denoted as $\Delta^k(t)$, which satisfies:
$$
|\Delta^k(t)|=10^{t-k}
$$
where $|\Delta^k(t)|$ represents the number of elements of set $\Delta^k(t)$.
Therefore, the local structure of the network $G(t)$ is similar to the overall structure of the network, that is, the network has a global self-similar property.

It is worth noting that network $G(t)$ can also be constructed in other iterative ways, that is, in each minimum unit $X^i(0)$ $(1\le i\le 10^{t-1})$ of network $G(t-1)$, the midpoints of the four faces and the third points of the six edges of $X^i(0)$ are defined as newly generated nodes of $G(t)$, and the edges between nodes (as shown in Fig \ref{fig1}) are defined as newly generated edges of $G(t)$. This iterative process is equivalent to replacing each of the minimum units $X^i(0)$ in network $G(t-1)$ with $X^i(1)$ and then $G(t)$ is constructed. In this iteration method, it can be found that the nodes in $G(t-1)$ are still retained after one iteration, and the newly generated nodes will appear inside each $X^i(0)$. Therefore, as shown in Fig \ref{fig1}, the nodes in the network $G(t)$ are marked with corresponding numbers. Next, in the network $G(t)$,  we define the set of all nodes as $V(t)$, and the set of all edges as $E(t)$. Among them, the set of newly generated nodes in the $t$-th iteration is denoted as $\bar{V}(t)$, which satisfies:
$$
\bar{V}(t)=\bigcup_{i=1}^{10^{t-1}}\bar{V}^{i}(t),
$$
where $\bar{V}^{i}(t)$ represent the set of newly generated nodes in $X^i(0)$ in $G(t-1)$. Therefore, we can naturally get the following relationship:
$$
V(t)=V(t-1)\cup\bar{V}(t)=V(t-1)\cup\bar{V}^{1}(t)\cup\bar{V}^{2}(t)\cup\cdots\cup\bar{V}^{10^{t-1}}(t).
$$
In addition, we define the set of the four outermost nodes of $X^i(k)$ as $V^i(k,t)$ in network $G(t)$. According to the structure of the network $G(t)$, the following relationship can be obtained:
$$
V(t-1)=\bigcup_{i=1}^{10^{t-1}}V^i(1,t).
$$
Since the trap node is set at node $1$, let $V'(0)=V(0)/\{1\}=\{2,3,4\}$.

Then, we define the number of nodes as $N(t)$ and the number of edges as $M(t)$ in the network $G(t)$, which satisfies the following equation:
\begin{align}
    M(t)&=6 \cdot |\Delta(t)|=6 \cdot 10^{t},\label{2.1}\\
    N(t)&=N(t-1)+|\Delta(t)| \cdot 16=4+\frac{16}{9}\left(10^{t}-1\right).\label{2.2}
\end{align}
In addition, the degree of node $i\in V(t)$ in the network $G(t)$, that is, the number of edges connected to the node $i$, is denoted as $d_i(t)$.
Observing Fig \ref{fig1}, we can find that the degree distribution of network $G(t)$ is quite uniform, where the degree of the three outermost nodes $1$, $2$, and $3$ is $3$, and the degree of the remaining nodes is $6$ or $9$. And the degree of the node does not change with the occurrence of iterations.

So far we have given the construction method of the $3$-dimensional $3$-level Sierpinski gasket network $G(t)$ and some basic properties.

\section{Mean Global First Passage Time on $G(t)$}

In this section, we will discuss the mean global first passage time (MGFPT) on the network $G(t)$ when the trap node is set to node $1$.
First, some basic definitions of random walk will be given.
For nodes $\forall i,j\in V(t)$, the first passage time (FPT) and mean first passage time (MFPT) from node $i$ to node $j$ are denoted as $F_{i,j}(t)$ and $T_{i,j}(t)$, where the target node $j$ can also be replaced with a node set, for example, $F_{i,V(0)}(t)$ ($T_{i,V(0)}(t)$) is the FPT (MFPT) from node $i$ to node set $V(0)$. When the node $i$ is not a trap node, the mean captured time (MCT) of node $i$ is equivalent to the MFPT from node $i$ to the trap node, denoted as $T_i(t)$, where the trap node defaults to node $1$, and when the trap node changes, we will make an explanation and give a new mark. 
In particular, $F_{i,i}(t)$ and $T_{i,i}(t)$ represent the first return time (FRT) and mean first return time (MFRT), but the MCT of $1$ is stipulated to satisfy $T_1(t)=0$.

After giving some basic definitions of random walks on the network $G(t)$, the mean global first passage time (MGFPT) of trap node $1$, recorded as $\langle GT\rangle(t)$, is defined as:
\begin{align}
    \langle GT(t)\rangle=\sum_{i=1}^{N(t)} \frac{d_{i}(t)}{2 M(t)} T_{i,1}(t)=\sum_{i \in V(t)/\{1\}} \frac{d_{i}(t)}{2 M(t)} T_{i}(t)+\frac{d_1(t)}{2 M(t)}T_{1,1}(t).\label{3.1}
\end{align}
Naturally, the global first pass time (GFPT) is recorded as $GT(t)$, and only needs to be changed from $T_{i,1}(t)$ to $F_{i,1}(t)$ in Eq(\ref{3.1}). In addition, the return time of node $1$ is denoted as $R(t)$. The difference between it and the $F_{i,i}(t)$ is that at time $R(t)$, the walker returns to the initial node $1$, but it is not necessarily the first time to return to this node.

In this section, the probability generating function (PGF) will occupy an extremely critical position in the MGFPT solution process. Here, the PGF of $F_{1,V'(0)}(t)$, $F_{1,1}(t)$, $R(t)$ and $GT(t)$ are respectively denoted as $\Phi_FPT(t)$, $\Phi_FRT(t)$, $\Phi_{RT}(t)$, $\Phi_{GT}(t)$, which are defined as\cite{gut2013probability}
\begin{align*}
    \Phi_{FPT}(t,z)&=\sum_{k=0}^{+\infty} z^{k} P\left\{F_{i,V'(0)}(t)=k\right\};~
    \Phi_{FRT}(t, z)=\sum_{k=0}^{+\infty} z^{k} P\left\{F_{1,1}(t)=k\right\}\\
    \Phi_{R T}(t, z)&=\sum_{k=0}^{+\infty} z^{k} P\left(R(t)=k\right);~
    \Phi_{GT}(t, z)=\sum_{k=0}^{+\infty} z^{k} P\left\{GT(t)=k\right\}
\end{align*}
where $P\left\{F_{i,j}(t)=k\right\}$ represents the probability of $F_{i,j}(t)=k$. Then, for any $t>0$, it is known that\cite{hwang2012first}
\begin{equation}\label{3.2}
\Phi_{\mathrm{FRT}}(t, z)=1-\frac{1}{\Phi_{\mathrm{RT}}(t, z)}
\end{equation}
and
\begin{equation}\label{3.3}
\Phi_{\mathrm{GT}}(t, z)=\frac{z}{1-z} \cdot \frac{d_{1}(t)}{2 M(t)} \cdot \frac{1}{\Phi_{\mathrm{RT}}(t, z)}.
\end{equation}

Through the method of path segmentation, the following relationships can be easily proved\cite{peng2018moments}:
\begin{equation}\label{3.4}
\Phi_{\mathrm{FPT}}(t, z)=\left.\Phi_{\mathrm{FPT}}(1, x)\right|_{x=\Phi_{\mathrm{FPT}}(t-1, z)}
\end{equation}
and
\begin{equation}\label{3.5}
\Phi_{\mathrm{RT}}(t, z)=\frac{\Phi_{\mathrm{RT}}(t-1, z)}{\phi\left(\Phi_{\mathrm{FPT}}(t-1, z)\right)}
\end{equation}
where $\phi(z)$ satisfies that
$$
\phi(z) \equiv \frac{\Phi_{\mathrm{RT}}(0, z)}{\Phi_{\mathrm{RT}}(1, z)}.
$$
Then, combining Eq.(\ref{3.2}), Eq.(\ref{3.3}), Eq.(\ref{3.4}) and Eq.(\ref{3.5}), we can obtain:
\begin{align}
\Phi_{\mathrm{FRT}}(t, z) &=1-\phi\left(\Phi_{\mathrm{FPT}}(t-1, z)\right)\left(1-\Phi_{\mathrm{FRT}}(t-1, z)\right)\label{3.6} \\
\Phi_{\mathrm{GT}}(t, z) &=\frac{1}{10} \phi\left(\Phi_{\mathrm{FPT}}(t-1, z)\right) \Phi_{\mathrm{GFPT}}(t-1, z).\label{3.7}
\end{align}

By the property of the PGF, $T_{1, V'(0)}(t)$,  $T_{1,1}(t)$ and $\langle GT(t)\rangle$ will satisfy the following relationships:
\begin{align}
    T_{1, V'(0)}(t)&=\left.\frac{\partial}{\partial z} \Phi_{\mathrm{FPT}}(t, z)\right|_{z=1}=T_{1, V'(0)}(1)T_{1, V'(0)}(t-1)=\lambda^t\label{3.8}\\
    T_{1,1}(t)&=\left.\frac{\partial}{\partial z} \Phi_{F R T}(t, z)\right|_{z=1}=\phi(1) T_{1,1}(t-1)=T_{1,1}(0)\cdot\rho_0^t\label{3.9}\\
    \left\langle GT(t)\right\rangle&= \left.\frac{\partial}{\partial z} \Phi_{GT}(t, z)\right|_{z=1}=\frac{1}{10}\left[\rho_0\left\langle GT(t-1)\right\rangle+\rho_{1}T_{1, V'(0)}(t-1)\right]\label{3.10}
\end{align}
where $\lambda=T_{1, V'(0)}(1)$, $\rho_0=\phi(1)$ and $\rho_1=\left.\frac{\partial\phi(x)}{\partial x}\right|_{x=1}$. Therefore, in order to obtain the analytical expression of $\langle GT(t)\rangle$, the unknown quantities $\lambda$, $\rho_0$ and $\rho_1$ must be solved.

For the above goals, we need to introduce matrix algorithms. Let the transition probability matrices of the networks $G(0)$ and $G(1)$ be $M_0$ and $M_1$ respectively, which can be expressed as:

\begin{align*}
    M_0=\left(\begin{array}{cccc}0 & \frac{1}{3} & \frac{1}{3} & \frac{1}{3} \\ \frac{1}{3} & 0 & \frac{1}{3} & \frac{1}{3} \\ \frac{1}{3} & \frac{1}{3} & 0 & \frac{1}{3} \\ \frac{1}{3} & \frac{1}{3} & \frac{1}{3} & 0\end{array}\right)
    \footnotesize
    \setlength{\arraycolsep}{2pt}
    M_1=\left(\begin{array}{cccccccccccccccccccc}
    0&\frac13&\frac13&\frac13&0&0&0&0&0&0&0&0&0&0&0&0&0&0&0&0\\
 \frac16&0&\frac16&\frac16&\frac16&\frac16&0&\frac16&0&0&0&0&0&0&0&0&0&0&0&0\\
 \frac16&\frac16&0&\frac16&0&\frac16&\frac16&0&\frac16&0&0&0&0&0&0&0&0&0&0&0\\
 \frac16&\frac16&\frac16&0&0&0&0&\frac16&\frac16&\frac16&0&0&0&0&0&0&0&0&0&0\\
 0&\frac16&0&0&0&\frac16&0&\frac16&0&0&\frac16&\frac16&0&0&\frac16&0&0&0&0&0\\
 0&\frac19&\frac19&0&\frac19&0&\frac19&\frac19&\frac19&0&0&\frac19&\frac19&0&0&\frac19&0&0&0&0\\
 0&0&\frac16&0&0&\frac16&0&0&\frac16&0&0&0&\frac16&\frac16&0&0&\frac16&0&0&0\\
 0&\frac19&0&\frac19&\frac19&\frac19&0&0&\frac19&\frac19&0&0&0&0&\frac19&\frac19&0&\frac19&0&0\\
 0&0&\frac19&\frac19&0&\frac19&\frac19&\frac19&0&\frac19&0&0&0&0&0&\frac19&\frac19&0&\frac19&0\\
 0&0&0&\frac16&0&0&0&\frac16&\frac16&0&0&0&0&0&0&0&0&\frac16&\frac16&\frac16\\
 0&0&0&0&\frac13&0&0&0&0&0&0&\frac13&0&0&\frac13&0&0&0&0&0\\
 0&0&0&0&\frac16&\frac16&0&0&0&0&\frac16&0&\frac16&0&\frac16&\frac16&0&0&0&0\\
 0&0&0&0&0&\frac16&\frac16&0&0&0&0&\frac16&0&\frac16&0&\frac16&\frac16&0&0&0\\
 0&0&0&0&0&0&\frac13&0&0&0&0&0&\frac13&0&0&0&\frac13&0&0&0\\
 0&0&0&0&\frac16&0&0&\frac16&0&0&\frac16&\frac16&0&0&0&\frac16&0&\frac16&0&0\\
 0&0&0&0&0&\frac19&0&\frac19&\frac19&0&0&\frac19&\frac19&0&\frac19&0&\frac19&\frac19&\frac19&0\\
 0&0&0&0&0&0&\frac16&0&\frac16&0&0&0&\frac16&\frac16&0&\frac16&0&0&\frac16&0\\
 0&0&0&0&0&0&0&\frac16&0&\frac16&0&0&0&0&\frac16&\frac16&0&0&\frac16&\frac16\\
 0&0&0&0&0&0&0&0&\frac16&\frac16&0&0&0&0&0&\frac16&\frac16&\frac16&0&\frac16\\
 0&0&0&0&0&0&0&0&0&\frac13&0&0&0&0&0&0&0&\frac13&\frac13&0
    \end{array}\right)
\end{align*}
From the relationship between the transition probability matrix of the network and the property of random walks, the following matrix algorithms can be easily proved\cite{zhang2021mean}:
\begin{enumerate}[(a)]
\item Let
$$
T_{V'(0)}=(T_{1,V'(0)}(1),T_{5,V'(0)}(1),T_{6,V'(0)}(1),\ldots,T_{20,V'(0)}(1))^{T}
$$
Denote the matrix obtained by deleting the rows and columns corresponding to nodes $2$, $3$, and $4$ in matrix $M_1$ as $M'_1$.
Then, the following relationship is naturally established:
$$
T_{V'(0)}=M'_1 T_{V'(0)}+e_{17} \Rightarrow T_{V'(0)}=(I_{17}-M'_1)^{-1} e_{17}
$$
where $e_{17}$ represents a column vector with $17$ elements all being $1$, and $I_{17}$ represents an identity matrix with a rank of $17$. By solving the vector $T_{V'(0)}$, we can naturally obtain the coefficient $\lambda=\frac{160}{9}$.

\item Let
\begin{align}\label{3.11}
    \Psi(k,z)=\sum_{n=0}^{+\infty}(z M_k)^{n}=(I_{N(k)}-z M_k)^{-1}=\left(\Psi_{i,j}(k,z)\right)_{N(k)\times N(k)}
\end{align}
where $k=0$ or $1$. It can be proved that $\Psi_{i,j}(k,z)$ is the probability generating function of the passage time from node $i$ to node $j$, and in particular $\Psi_{1,1}(k,z)=\Phi_{\mathrm{RT}}(k, z)$. 
Therefore, substituting the matrices $M_0$ and $M_1$ into Eq.(\ref{3.11}), we can obtain:
\begin{align*}
    \Phi_{\mathrm{RT}}(0, z)&=\frac{2z - 3}{z^2 + 2z - 3}\\
    \Phi_{\mathrm{RT}}(1, z)&=-\frac{z^4 + 11z^3 - 195z^2 + 504z - 324}{z^5 - 17z^4 + 55z^3 + 141z^2 - 504z + 324}
\end{align*}
Therefore, 
$$
\phi(z)=-\frac{2z^4 - 41z^3 + 249z^2 - 504z + 324}{z^4 + 11z^3 - 195z^2 + 504z - 324}.
$$
Then, we can naturally get $\rho_0=10$ $\rho_1=463$.
Similarly, using the matrix algorithm, we can also get the following initial conditions in the initial network $G(0)$:
$$
T_{1,1}(0)=4,~~\langle GT(0)\rangle=\frac{13}{4}.
$$
\end{enumerate}

\begin{figure}[t]
\centering
\includegraphics[scale=0.7]{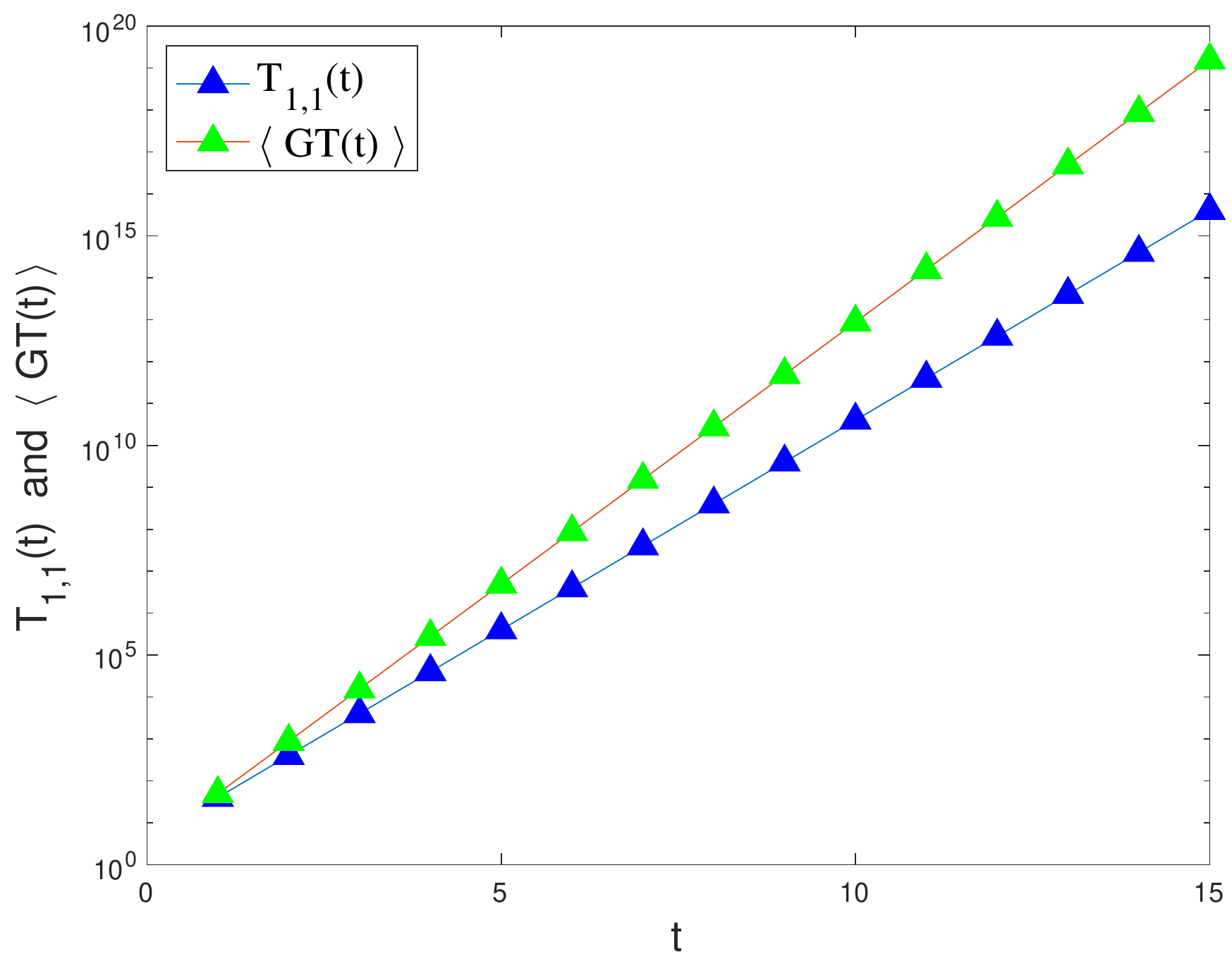}
\caption{Numerical simulation diagram of $T_{1,1}(t)$ and $\left\langle GT(t)\right\rangle$ with respect to the number of iterations $t$.}
\label{fig2}
\end{figure}

Therefore, only need to substitute $\lambda$, $\rho_0$ and $\rho_1$ into Eq.(\ref{3.8}) and Eq.(\ref{3.9}), we can naturally get:
\begin{align}
    T_{1, V'(0)}(t)&=\lambda^t=\left(\frac{160}{9}\right)^t\label{3.12}\\
    T_{1,1}(t)&=T_{1,1}(0)\cdot\rho_0^t=4\cdot10^t\label{3.13}   
\end{align}
In addition,
\begin{align}
    \left\langle GT(t)\right\rangle &=\frac{1}{10}\left[\rho_0\left\langle GT(t-1)\right\rangle+\rho_{1}T_{1, V'(0)}(t-1)\right]\nonumber\\
    &=\frac{463}{10} \cdot\left(\frac{160}{9}\right)^{t-1}+\langle G T(t-1)\rangle\nonumber\\
    &=\langle GT(0)\rangle+\frac{463}{10} \sum_{i=0}^{t-1}\left(\frac{160}{9}\right)^{i}\nonumber\\
    &=\frac{1940}{703} \cdot\left(\frac{160}{9}\right)^{t}+\frac{383}{781}\label{3.14}
\end{align}
Finally, we performed a numerical simulation of $T_{1,1}(t)$ and $\left\langle GT(t)\right\rangle$, and the results are shown in Fig\ref{fig2}.

\section{Average Trapping Time on $G(t)$}

In this section, we will calculate the average trapping time (ATT) for the network $G(t)$, in which MGFPT will play a key role. First of all, the sum of mean captured times on network $G(t)$, denoted by $S_k(t)$, is defined as:
$$
S_k(t)=\sum_{i\in V(k)} T_{i}(t),
$$
where $k\le t$.
Then, the ATT of trap node $1$ on network $G(t)$, denoted by $\langle A T(t)\rangle$, is defined as follows:
\begin{eqnarray}\label{4.1}
\langle A T(t)\rangle=\frac{1}{N(t)-1} S_t(t) =\frac{1}{N(t)-1}\sum_{i\in V(t)} T_{i}(t).
\end{eqnarray}
It can be easily proved that $S_0(0)=9$.

In order to solve $S_t(t)$, the sum of the MCT for the node set $\bar{V}(t)$ is presented and defined as:
$$
\bar{S}(t)=\sum_{i \in \bar{V}(t)} T_{i}(t).
$$
Therefore
\begin{align}\label{4.0}
    S_t(t)=S_{t-1}(t)+\bar{S}(t).
\end{align}

Then, $S_{t-1}(t)$ and $S_{t-1}(t)$ are considered separately.
\begin{enumerate}[(a)]
\item Based on Eq(\ref{3.8}) and the properity of random walks on the network, it can be found that
$$
T_{i,j}(t)=\lambda \cdot T_{i,j}(t-1),
$$
where $i,j\in V(t-1)$. From this, we can deduce:
\begin{eqnarray}\label{4.2}
S_{t-1}(t)=\lambda \cdot S_{t-1}(t-1)=\frac{160}{9}\cdot S_{t-1}(t-1).
\end{eqnarray}
\item For any area $X^i(1)$ in the network $G(t)$, the path of its internal node $k\in \bar{V}^i(t)$ to the trap node $1$ can be divided into two sections:

(1) Starting at node $k$, the walker arrives at one of the four outermost nodes $l\in V^i(1,t)$ for the first time after a random walk. It is worth noting that since the four nodes in $V^i(1,t)$ are symmetrical in region $X^i(1)$, starting from the $16$ nodes in $\bar{V}^i(t)$, the four nodes in $ V^i(1,t)$ will receive $4$ walkers on average. In addition, the number of regions $X^i(1)$ ($1\le i\le10^{t-1}$) connected to node $l$ is $\frac{d_l(t)}{3}$.

(2) The walker then starts from node $l\in V^i(1,t)$ and finally reaches the trap node $1$.

Moreover, based on the matrix algorithm, it can be obtained that:
$$
\sigma=\sum_{j\in \bar{V}^i(t)}T_{j, V^i(1,t)}(t)=150.
$$
According to the above analysis, we can obtain:
\begin{align*}
    \bar{S}(t)&=\sum_{i \in \bar{V}(t)} T_{i}(t)\nonumber\\
    &=\sigma\cdot|\Delta^1(t)|+4\cdot \sum_{j\in V(t-1)} \frac{d_j(t)}{3}T_j(t)\nonumber\\
    &=\sigma\cdot|\Delta^1(t)|+\frac{4}{3}\lambda\cdot  \sum_{j\in V(t-1)} d_j(t-1)T_j(t-1)\
\end{align*}
From the definition of MGFPT, the following relationship can be obtained:
$$
\begin{aligned}
&\sum_{j\in V(t-1)} d_j(t-1)T_j(t-1)\\
=&\left[\langle G T(t-1)\rangle-\frac{d_1(t-1)}{2 M(t-1)}T_{1,1}(t-1)\right] \cdot 2 M(t-1)\\
=&[\langle G T(t-1)\rangle-1] \cdot 2 M(t-1)
\end{aligned}
$$
Therefore, 
\begin{align}\label{4.3}
\bar{S}(t)=150 \cdot\left(10^{t-1}\right)+\frac{4}{9} \cdot 1280 \cdot 10^{t-1}\left[\frac{1940}{703} \cdot\left(\frac{160}{9}\right)^{t-1}-\frac{1164}{781}\right]
\end{align}
\end{enumerate}

Substituting Eq(\ref{4.2}) and Eq(\ref{4.3}) into Eq(\ref{4.0}), we can obtain:
\begin{align}\label{4.4}
{S}_{t}(t)
&=\frac{160}{9} S_{t-1}(t-1)-\frac{16051}{23} \cdot 10^{t-1}+\frac{19501}{23} \cdot 10^{t-1} \cdot\left(\frac{160}{9}\right)^{t-1}\nonumber\\
&=S_0(0)\left(\frac{160}{9}\right)^{t}-\frac{16051}{23}\sum_{i=0}^{t-1}10^{i}\left(\frac{160}{9}\right)^{t-1-i}+\frac{19501}{23}\left(\frac{160}{9}\right)^{t-1} \sum_{i=0}^{t-1} 10^{i}\nonumber\\
&=\frac{1293}{244}\left(\frac{160}{9}\right)^{t} \cdot 10^{t}-\frac{10237}{119} \cdot\left(\frac{160}{9}\right)^{t}+\frac{6550}{73} \cdot 10^{t}
\end{align}
Substituting Eq(\ref{4.4}) into Eq(\ref{4.1}), the analytical expression of the ATT can be obtained:
\begin{align}\label{4.5}
    \langle A T(t)\rangle=\frac{9\left[\frac{1293}{244}\left(\frac{160}{9}\right)^{t} \cdot 10^{t}-\frac{10237}{119} \cdot\left(\frac{160}{9}\right)^{t}+\frac{6550}{73} \cdot 10^{t}\right]}{27+16\left(10^{t}-1\right)}.
\end{align}

\section{Joint Network model of $G(t)$}
In the previous section, we finally obtained the ATT when the outermost node $1$ is a trap node. Therefore, a natural question is how to solve the analytical expression of ATT on the network $G(t)$ when the trap node is not $1$, and there are even multiple trap nodes. In this section, we will introduce a joint network model and solve the above problems based on this model.

First, we define the mode variable $\Omega$ to be a network formed by splicing $m$ tetrahedrons, in which the splicing method is defined as point-to-point coincidence and non-coincidence of edges. 
Then, the joint network model whose mode variable is $\Omega$, denoted as $JG(\Omega,t)$, is defined as the network model formed by replacing the m tetrahedrons in $\Omega$ with the network $G(t)$. Obviously, the joint network model $JG(\Omega,t)$ is an extension of $G(t)$, because when $\Omega=G(1)$, $JG(G(1),t-1)=G(t)$.

In network $JG(\Omega,t)$, let the set of nodes be $V(\Omega,t)$, the set of edges be $E(\Omega,t)$, the number of nodes is $N(\Omega,t)$, and the number of edges is $M(\Omega,t)$. In addition, $JG(\Omega,t)$ is formed by splicing $m$ self-similar regions $X^i(t)$ ($1\le i\le m$), so the set of all nodes inside each $X^i(t)$ (not including the four outermost nodes of $X^i(t)$) is denoted as $\bar{V}^i(\Omega,t)$. And the set of the four outermost nodes of $X^i(t)$ is $\hat{V}^i(\Omega,t)$. Therefore, we have:
$$
V(\Omega,t)=\bigcup_{i=1}^m \hat{V}^i(\Omega,t)\cup\bigcup_{i=1}^m \bar{V}^i(\Omega,t)
\triangleq V(\Omega,0) \cup \bar{V}(\Omega,t).
$$
Let the set of trap nodes be $\theta$, which satisfies $\theta\in V(\Omega,0)$, and the number of nodes in $\theta$ is $n$.

Then, in the network $JG(\Omega,t)$, define the MFPT from node $i$ to node $j$ as $T_{i,j}(\Omega,t)$, and the average capture time from node $i$ to trap node set $\theta$ as $T_{i}(\Omega,t)$. Let
$$
S'(t)=\sum_{i\in V(t)/V(0)}T_{i,V(0)}(t)~~~\textrm{and} ~~~
S(\Omega,t)=\sum_{i\in V(t)}T_{i}(\Omega,t).
$$
Therefore, the ATT on $JG(\Omega,t)$, denoted as $\langle AT(\Omega,t)\rangle$, can be expressed as:
\begin{align}\label{5.1}
    \langle AT(\Omega,t)\rangle=\frac{S(\Omega,t)}{N(\Omega,t)-n}.
\end{align}

On network model $G(t)$, from Eq.(\ref{3.12}), we know that:
$$
T_{1,2}(t)=T_{1,V'(0)}(t)+\frac{2}{3} T_{1,2}(t)
\Rightarrow T_{1,2}(t)=3 T_{1,V'(0)}(t)=3 \cdot\left(\frac{160}{9}\right)^{t}
$$
Then, the following relationship established:

\begin{align*}
S(t) &=S'(t)+\left[\frac{3}{4}(N(t)-4)+3\right] T_{1, 2}(t) \nonumber\\
&=S'(t)+\frac{3}{4} N(t) T_{1,2}(t) \nonumber\\
\Rightarrow
S'(t) &=\frac{317}{244}\left(\frac{160}{9}\right)^{t} \cdot 10^{t}-\frac{10832}{119}\left(\frac{160}{9}\right)^{t}+\frac{6550}{73} 10^{t}
\end{align*}

Based on the above-mentioned random walk properties on the network $G(t)$, we can make the following analysis: 
\begin{enumerate}[(a)]
\item In network $JG(\Omega,t)$, the MCT of node $\forall i\in V(\Omega,t)$ to the set of trap nodes satisfies the following relationship:
$$
T_i(\Omega,t)=\lambda^t\cdot T_i(\Omega,0).
$$
Therefore, we can naturally obtain:
\begin{equation}\label{5.2}
\sum_{i \in V(\Omega,0)} T_{i}(\Omega,t)
=\lambda^t \sum_{i \in V(\Omega,0)} T_{i}(\Omega, 0)
\triangleq T(\Omega) \cdot \lambda^t
\end{equation}
Here, $T(\Omega)$ is only determined by the mode variable $\Omega$ and the set of trap nodes $\theta$, and can be obtained by matrix algorithm.

\item For node $\forall i\in \bar{V}^k(\Omega,t)$, the path of walker starting from this node and finally being captured by the trap node set $\theta$ can be divided into two sections:

(1) The particle starts from node $i$ and reaches any node in set $\hat{V}^k(\Omega,t)$ for the first time; 

(2) Then starts from this node and finally reaches the trap node set $\theta$.

From this, the following relationship can be obtained:
\begin{align}\label{5.3}
    \sum_{i \in\bar{V}(\Omega,t)}T_i(\Omega,t)&=\sum_{k=1}^{m}\sum_{i \in\bar{V}^k(\Omega,t)}T_i(\Omega,t)\nonumber\\
    &= \sum_{k=1}^{m}\left[S'(t)+\frac{N(t)-4}{4}\sum_{j \in\hat{V}^k(\Omega,t)}T_j(\Omega,t)\right]\nonumber\\ 
    &=m\cdot S'(t)+ \frac{N(t)-4}{12}\cdot \lambda^t\sum_{j \in\hat{V}(\Omega,0)} d_i(0) T_j(\Omega,0)\nonumber\\
    &\triangleq m\cdot S'(t)+ \frac{N(t)-4}{12}\cdot \lambda^t\cdot T_d(\Omega).
\end{align}
Similarly, $T_d(\Omega)$ is only determined by the mode variable $\Omega$ and the set of trap nodes $\theta$, and can be obtained by matrix algorithm.

\end{enumerate}

Based on Eq(\ref{5.2}) and Eq(\ref{5.3}), we can obtain:
\begin{align}\label{5.4}
    S(\Omega,t)=&\sum_{i \in V(\Omega,0)} T_{i}(\Omega, t)+\sum_{j \in \bar{V}(\Omega,t)} T_{j}(\Omega, t)\nonumber\\
    =&T(\Omega)\cdot\lambda^t+m\cdot S'(t)+ \frac{N(t)-4}{12}\cdot \lambda^t\cdot T_d(\Omega)\nonumber\\
    =&\left(\frac{317}{244}m+\frac{4}{27} T_d(\Omega)\right)\left(\frac{160}{9}\right)^{t} \cdot 10^{t}\nonumber\\ &+\left(T(\Omega)-\frac{10832}{119}m-\frac{4}{27} T_d(\Omega)\right)\left(\frac{160}{9}\right)^{t}\nonumber\\
    &+\frac{6550}{73}m \cdot 10^{t}
\end{align}
Finally, just substituting Eq.(\ref{5.4}) into Eq.(\ref{5.1}), we can get the analytical expression of ATT with the trap node set $\theta$ on the network $J(\Omega,t)$.

Then, some simulation examples are given to illustrate the theoretical results.
Here, we let $\theta=\{8, 9, 10\}$ be a trap nodes set and calculate the ATT on network $G(t)$. 

Let the mode variable $\Omega=G(1)$, we have $G(t)=JG(G(1),t-1)$, where the transition probability matrix of $G(1)$ is $M_1$ presented in the Section 3.
In order to find the MCT for the non-trap node to reach the trap node set on $G(1)$, we only need to repeat the process in 3(a). If the matrix after deleting the rows and columns corresponding to trap nodes $8,9,10$ in $M_1$ is denoted as $M_1^{\prime\prime}$, the vector 
$$
T_{\theta}=(T_{1,\theta}(1),\ldots,T_{7,\theta}(1),T_{11,\theta}(1),\ldots,T_{20,\theta}(1))^T
$$
satisfies the following relationship:
$$
T_{\theta}=M_1^{\prime\prime} T_{\theta}+e_{17} \Rightarrow T_{\theta}=(I_{17}-M_1^{\prime\prime})^{-1} e_{17}.
$$
Based on Matlab, we can naturally obtain
\begin{align*}
    T_{\theta}=&\Big(
    \frac{1598}{253},
    \frac{2754}{581} ,
    \frac{2777}{653}   ,
    \frac{2213}{287}  ,
    \frac{1390}{253} ,  
    \frac{1821}{388}  , 
    \frac{1665}{289}   ,
    \frac{6029}{1314}  ,\\
    &\frac{3963}{707}   ,
    \frac{3817}{591}   ,
    \frac{2879}{535}   ,
    \frac{2651}{516}   ,
    \frac{2575}{441}   ,
    \frac{1747}{292}   ,
    \frac{633}{137}   ,
    \frac{8152}{1113}  ,
    \frac{4451}{701}
    \Big)^T.
\end{align*}
Then, we have
$$
T(G(1))=\frac{16941}{176}~~~\textrm{and}~~~
T_d(G(1))=\frac{130339}{234}.
$$
Substituting the above parameters $T(G(1))$ and $T_d(G(1))$ into Eq(\ref{5.4}), the sum of MCT on $G(t)$ with trap nodes set $\theta$ satisfy: 
\begin{align*}
    S(G(1),t-1)=
    \frac{13085}{137}\left(\frac{160}{9}\right)^{t-1} \!\!\cdot 10^{t-1} -\frac{29585}{33}\left(\frac{160}{9}\right)^{t-1}\!\!+\frac{6550}{73} \cdot 10^{t}.
\end{align*}
Then, the analysis expression of ATT on $G(t)$ with trap nodes set $\theta$, denoted as $\langle A T(G(1), t-1)\rangle$ is 
\begin{align*}
    \langle A T(G(1), t-1)\rangle&=\frac{S(G(1),t-1)}{N(t)-3}\\
    &=\frac{9\left[\frac{13085}{137}\left(\frac{160}{9}\right)^{t-1} \!\!\cdot 10^{t-1} -\frac{29585}{33}\left(\frac{160}{9}\right)^{t-1}\!\!+\frac{6550}{73} \cdot 10^{t}\right]}{16\cdot 10^t-7}.
\end{align*}
In order to compare the relationship between the ATT when the trap node sets are different, we performed numerical simulations on the above results and the result when the trap node is $1$, as shown in Figure 3. From the comparison chart, it can be found that when the trap node set changes, the ATT will change significantly, but the trend of the number of iterations will not be changed.

\begin{figure}[t]
\centering
\includegraphics[scale=0.7]{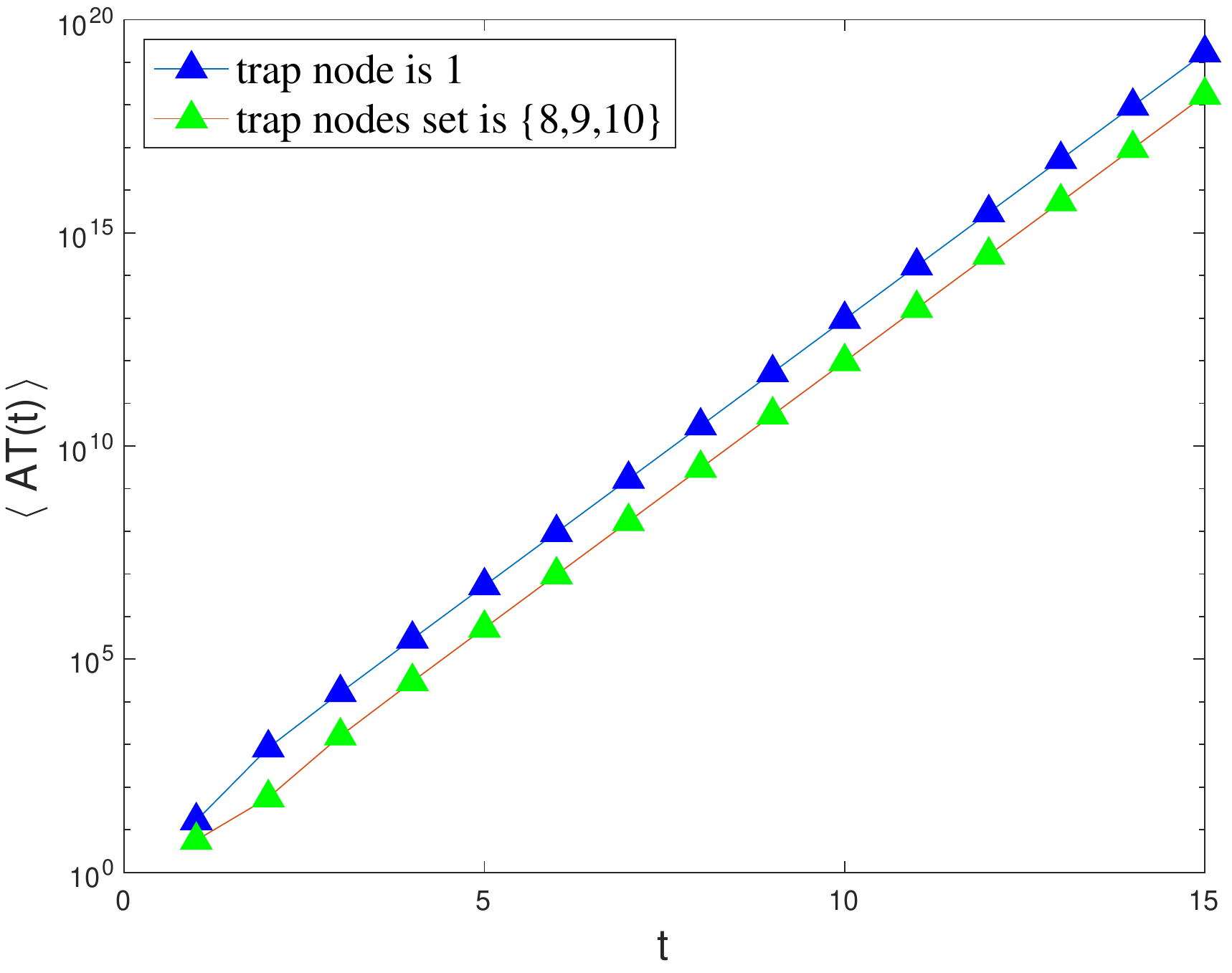}
\caption{The numerical simulation diagram of ATT with respect to the number of iterations $t$, in which the trap node sets are set to $\{1\}$ and $\{8,9,10\}$.}
\label{fig3}
\end{figure}

\section{Conclusion}
In this article, we first introduced the construction method and structural characteristics of the 3-dimensional 3-level Sierpinski gasket network $G(t)$ in the second 2. 
Then, in the section 3, based on the probability generating function, matrix algorithm and path segmentation method, the analytical expression of MGFPT on network $G(t)$ is solved. 
Furthermore, in the section 4, we use the analytical expression of MGFPT and the random walk feature to obtain the analytical expression of the ATT of the trap node $1$.
Finally, in order to extend the ATT in the setting of trap nodes, we introduced the joint network model $JG(\Omega,t)$ in the section 5. Based on the matrix algorithm, the analytical expression of the ATT with the trap node set $\theta$ on the network $JG(\Omega,t)$ is obtained, where the trap node set can only be set on the nodes in the model variable $\Omega$. Through numerical simulation, we also verified that the change of trap node set will significantly affect the ATT on the network. 

It is worth noting that in this article, although we only discuss the random walk problem on the 3-dimensional 3-level Sierpinski Sierpinski gasket network, the method used can naturally be extended to any $n$-dimensional $m$-level network, where $n,m\ge 2$.

\section*{Acknowledgments}
This work was supported by the National Natural Science Foundation of China (Grant No. 12026214).

\bibliographystyle{unsrt}

\bibliography{ref}

\end{document}